\documentclass[11pt,a4paper,twoside]{article} 
\usepackage{amsmath}
\usepackage{amsfonts}
\usepackage[normalem]{ulem}
\usepackage{color}
\usepackage{rotating}
\usepackage{graphicx}
\usepackage[T1]{fontenc} 
\usepackage{bm}
\usepackage{hyperref}
\usepackage{version}

\numberwithin{equation}{section}

\begin{document}

\noindent
{\bf
{\Large Topologically protected duality on the boundary of Maxwell-BF theory \\

}} 

\vspace{.5cm}
\hrule

\vspace{1cm}

\noindent
{\large{Alberto Blasi$^{a,}$\footnote{\tt alberto.blasi@ge.infn.it },
Nicola Maggiore$^{a,b,}$\footnote{\tt nicola.maggiore@ge.infn.it },
\\[1cm]}}

\noindent
{{}$^a$ Dipartimento di Fisica, Universit\`a di Genova,\\
via Dodecaneso 33, I-16146, Genova, Italy\\
{}$^b$ I.N.F.N. - Sezione di Genova}
\vspace{1cm}

{\tt Abstract~:}\\

The Maxwell-BF theory with a single-sided planar boundary is considered in Euclidean four dimensional spacetime. The presence of a boundary breaks the Ward identities which describe the gauge symmetries of the theory, and, using standard methods of quantum field theory, the most general boundary conditions and a nontrivial current algebra on the boundary are derived.  The electromagnetic structure which characterizes the boundary is used to identify the three dimensional degrees of freedom, which turn out to be formed by a scalar field and a vector field, related by a duality relation. The induced three dimensional theory shows a strong-weak coupling duality which separates different regimes described by different covariant actions. The role of the Maxwell term in the bulk action is discussed, together with the relevance of the topological nature of the bulk action for the boundary physics.

\vfill\noindent
{\footnotesize 
{\tt Keywords:}
Quantum Field Theory, Topological Quantum Field Theory, Duality in Gauge Field Theories,  Boundary Quantum Field Theory.
\\
{\tt PACS Nos:} 
03.70.+k	Theory of quantized fields;
11.10.-z	Field theory;
}

\newpage

\section{Introduction}\label{intro}

Topological field theories have been the subject of a thorough investigation in theoretical physics \cite{Witten:1982im,Witten:1988ze,Atiyah:1989vu}. The initial aim was to unveil to what extent they could give hints in better understanding gravity without matter \cite{Witten:1988hc}, but it was soon recognized that they also had and still have a different role if one adds a boundary \cite{Moore:1989yh,Cattaneo:2012zs}. Indeed it is the boundary which  plays a physical role and it is on the boundary that the local observables of a new and different physics live.
The introduction of a boundary in a field theory was first proposed by  Symanzik who introduced a separability ansatz in order to study the Casimir effect of two parallel plates \cite{Symanzik:1981wd}. The method proposed by Symanzik concerns a space divided into a left and a right hand side, and it has been applied to topological field theories of different types \cite{Amoretti:2012kb}, obtaining results particularly relevant for the theory of the  fractional quantum Hall effect \cite{Blasi:2008gt} and of the topological insulators in three and four spacetime dimensions \cite{Amoretti:2014iza}.  Later on  field theories with a single-sided boundary have also been considered, which also lead to interesting results.
The first and most studied model is the Chern-Simons theory in three dimensions \cite{Chern:1974ft,Blasi:1990pf}, soon after appeared the BF models \cite{Horowitz:1989ng}, the generalizations of the Chern-Simons model which can only live in an odd spacetime \cite{Karlhede:1989hz,Blasi:2005vf} .
The topological nature of the Chern-Simons  action is that it does not depend on the metric tensor and hence the energy momentum tensor vanishes; later on Witten  proposed another kind of topological theory where the action is not in the cohomology of the BRS operator and hence the theory has no physical observables \cite{Birmingham:1991ty}. 
The common denominator to all these models is that they are gauge field theories with the gauge field $A_\mu$ as the main actor and where the Maxwell term does not appear since it breaks the topological nature of the model, which seems to be at the basis of duality relations characterizing the boundary degrees of freedom of these models, whose bulk theories are purely topological \cite{Amoretti:2014kba}. Duality relations of this type are known \cite{Aratyn:1983bg,Aratyn:1984jz} to allow to extract fermionic degrees of freedom out of bosonic ones, in a way compatible with the existence of Hall or quantum spin states on the edge of higher dimensional bulk theories. 
It is natural to ask the question to what extent this is true, investigating whether fermionizing duality relations hold also on the boundary of non topological field theories, which would broaden the possible candidates for the theories of fractional quantum Hall effect and of topological insulators. 
Moreover, the Maxwell coupling is expected to be quite relevant in whatever physics may arise on the boundary and that is why we are studying models where the Maxwell term is included in the bulk action. This has been done for Chern-Simons theory with both double \cite{Blasi:2010gw} and single-sided \cite{Maggiore:2018bxr} boundary, with significantly different results. We note also that the introduction of a Maxwell term in the Chern-Simons action gives rise to topologically massive theories \cite{Deser:1981wh,Deser:1982vy} by means of a mechanism which cannot to be replicated in spacetime dimensions other than three. The question also arises in what sense a topological theory with a Maxwell coupling is still topological. We propose here the Maxwell-BF model  as a new kind of topological theory in four spacetime dimensions. The model does not fit into the known ``topological'' classes of quantum field theories since it does depend explicitly on the metric and the Maxwell term makes it cohomologically non trivial. Nevertheless the bulk theory has  no local observables due to the equations of motion which enforce the field strength $F_{\mu\nu}$ to vanish, and hence the gauge field $A_\mu$ is pure gauge. This model has never been considered before with a boundary and the question is not irrelevant. The fact that the physics of the model live on the boundary is neither intuitive nor immediately deducible as a simple exercise.\\

The paper is organized as follows. In Section 2 the model with planar boundary is introduced. The boundary conditions and the Ward identities, broken by the presence of the boundary, are derived. A kind of electromagnetic structure is found on the boundary, with Maxwell equations solved by potentials, which will play the role of degrees of freedom for the 3D theory. The identification of electric and magnetic fields makes possible a  physical interpretation of the role of the Maxwell term, as deformation of the magnetic field. In Section 3, following standard methods of quantum field theory, the algebra formed by the conserved electromagnetic currents is found, which heavily depends on the Maxwell term in the bulk action. The algebra, written in terms of the 3D potentials, allows for the construction, in Section 4, of the 3D theory, whose symmetries are identified. The holographic contact is realized by means of the equations of motion of the 3D theory, which are required to be compatible with the boundary conditions of the bulk 4D theory. The resulting equation is recognized to be the duality relation which characterizes the existence of fermionic degrees of freedom on the boundary and which therefore turns out not to be peculiar of purely topological bulk field theories only. Or, putting it in another way, the fact that the physical properties are the same in  the holographic theory whether the Maxwell term is present or not, clarifies the meaning of topological quantum field theories when a boundary is introduced. 


\section{The model: action, boundary conditions and Ward identities}\label{sect2}

The action of the 4D Maxwell-BF theory in euclidean spacetime 
is 
\begin{equation}
S_{bulk}=\int d^4x\, \theta(x_3) \left( 
k_1\,\epsilon_{\mu\nu\rho\sigma}F_{\mu\nu}B_{\rho\sigma} 
+ k_2\, F_{\mu\nu}F_{\mu\nu}\right)\ ,
\label{Sbulk}\end{equation}
where the presence of the step-function $\theta(x_3)$ restricts the model on the half-space $x_3\geq 0$, with planar boundary at $x_3=0$, $F_{\mu\nu}(x)$ is the electromagnetic tensor for the gauge field $A_\mu(x)$ and $B_{\mu\nu}(x)=-B_{\nu\mu}(x)$ is the rank-2 antisymmetric tensor of the 4D topological BF theory \cite{Blau:1989dh,Blau:1989bq}, and the canonical mass dimensions of the quantum fields are 
\begin{equation}
[A]=1 \ \ ;\ \  [B]=2\ .
\label{dimensions}\end{equation}
Finally, $k_1$ and $k_2$ are constants which could be reabsorbed by a redefinition of the fields, but which we prefer to keep, in order to be able to trace the contributions of the topological $BF$ and Maxwell $F^2$ term, respectively.

In absence of boundary, $i.e.$ without the $\theta$-function in \eqref{Sbulk}, the action of the Maxwell-BF theory is invariant under the following two symmetries: 
\begin{equation}
\begin{split}
\delta^{(1)} A_\mu &= \partial_\mu\Lambda \\
\delta^{(1)}B_{\mu\nu} &=0
\end{split}
\label{delta1}\end{equation}
and
\begin{equation}
\begin{split}
\delta^{(2)} A_\mu &= 0 \\
\delta^{(2)}B_{\mu\nu} &=\partial_\mu\zeta_\nu-\partial_\nu\zeta_\mu\ ,
\end{split}
\label{delta2}\end{equation}
where $\Lambda(x)$ and $\zeta_\mu(x)$ are gauge parameters. The presence of the boundary in $S_{bulk}$ breaks the $\delta^{(2)}$- invariance, preserving the usual gauge symmetry $\delta^{(1)}$:
\begin{equation}
\delta^{(1)}S_{bulk}=0\ \ ;\ \ \delta^{(2)}S_{bulk}=-4k_2\int d^4x \delta(x_3)\zeta_i\epsilon_{ijk}\partial_jA_k\ ,
\label{brokensym}\end{equation}
where latin indices run from 0 to 2: $i,j,...=\{0,1,2\}$.

The total action $S_{tot}$ is composed by four terms
\begin{equation}
S_{tot}=S_{bulk}+S_{gf}+S_{ext}+S_{bd}\ ,
\label{Stot}\end{equation}
where $S_{bulk}$ is given by \eqref{Sbulk}, $S_{gf}$ is the gauge fixing term
\begin{equation}
S_{gf}=\int d^4x\, \theta(x_3)\left(bA_3+d_iB_{3i}\right)\ ,
\label{Sgf}\end{equation}
and $b(x)$ and $d_i(x)$ are Lagrange multipliers implementing the gauge conditions
\begin{equation}
A_3=B_{3i}=0\ .
\label{gaugecond}\end{equation}
In $S_{ext}$ external sources $J_i(x)$ and $J_{ij}(x)$ are introduced 
\begin{equation}
S_{ext}=\int d^4x\, \theta(x_3)\left(J_iA_i+\frac{1}{2}J_{ij}B_{ij}\right)\ ,
\label{Sext}\end{equation}
by means of which the quantum fields surviving the gauge conditions \eqref{gaugecond} can be defined. 
Finally, $S_{bd}$ is the most general boundary  term defined on $x_3=0$ compatible with power counting
\begin{equation}
S_{bd}=\int d^4x\, \delta(x_3)\left(
a_1\epsilon_{ijk}A_i\partial_jA_k + a_2A_i\partial_3A_i +a_3\epsilon_{ijk}A_iB_{jk} + \frac{m}{2}A_iA_i
\right)\ ,
\label{Sbd}\end{equation}
where $a_i$ and $m$ are constant parameters and the canonical mass assignments \eqref{dimensions} and $[\delta]=1$ have been used.
Notice that 
\begin{equation}
\begin{split}
\int d^4x\, \theta(x_3) \epsilon_{\mu\nu\rho\sigma}F_{\mu\nu}F_{\rho\sigma} 
&= 
2\int d^4x\, \theta(x_3)\partial_\mu (\epsilon_{\mu\nu\rho\sigma}A_\nu F_{\rho\sigma})\\
&=-4\int d^4x\, \delta(x_3)\epsilon_{ijk}A_i\partial_jA_k\ ,
\end{split}\label{FFdual}\end{equation}
which justifies the fact that we did not introduce in $S_{bulk}$ \eqref{Sbulk} a term $F\tilde{F}$ in favor of its Chern-Simons-like boundary counterpart, identified by the constant $a_1$ in \eqref{Sbd}. In addition, we chose to keep 3D covariance on the boundary. A more general, non covariant boundary term could have been written \cite{Maggiore:2017vjf,Blasi:2015lrg,Blasi:2017pkk}.  \\

The equations of motion of the fields $A_i$ and $B_{ij}$ are
\begin{equation}
\begin{split}
\frac{\delta S_{tot}}{\delta A_i} &=
\theta(x_3)
[-2k_1\epsilon_{ijk}\partial_3B_{jk} - 4k_2\partial_3^2A_i-4k_2\partial_j^2A_i + 4k_2\partial_i\partial_jA_j +J_i]
\\
&+\delta(x_3)
[(a_3-2k_1)\epsilon_{ijk}B_{jk} + (a_2-4k_2) \partial_3A_i +2a_1\epsilon_{ijk}\partial_jA_k+mA_i]
=0
\end{split}
\label{eomA}\end{equation}
and
\begin{equation}
\frac{\delta S_{tot}}{\delta B_{ij}} =
\theta(x_3)
[4k_1\epsilon_{ijk}\partial_3A_k+ J_{ij}]
+\delta(x_3)
[2a_3\epsilon_{ijk}A_k]
=0\ .\label{eomB}
\end{equation}
From the equations of motion \eqref{eomA} and \eqref{eomB} and performing $\lim_{\epsilon\rightarrow 0} \int_{-\epsilon}^{+\epsilon} dx_3$, we get the boundary conditions
\begin{equation}
\left.(a_3-2k_1)\epsilon_{ijk}B_{jk} + (a_2-4k_2) \partial_3A_i +2a_1\epsilon_{ijk}\partial_jA_k+mA_i \right|_{x_3=0}= 0
\label{bcA} 
\end{equation}
\begin{equation}
\left.a_3A_i\right|_{x_3=0} = 0\ .
\label{bcB}
\end{equation}
The equations of motion lead also to the following Ward identities, broken by terms on the r.h.s. due to the presence of the boundary
\begin{equation}
\int_0^\infty dx_3\, \partial_iJ_i = 
-\partial_i\left.\left(2k_1\epsilon_{ijk}B_{jk}+4k_2\partial_3A_i\right)\right|_{x_3=0}
\label{wardA}\end{equation}
\begin{equation}
\int_0^\infty dx_3\, \partial_jJ_{ij}= 
(4k_1-2a_3)\left. \epsilon_{ijk}\partial_jA_k\right|_{x_3=0}\ ,
\label{wardB}\end{equation}
where, to obtain \eqref{wardA}, the boundary conditions \eqref{bcA} have been used. On the boundary $x_3=0$ and at vanishing external sources, $i.e.$ on the mass shell, the above Ward identities imply
\begin{equation}
\left.\partial_i\left(2k_1\epsilon_{ijk}B_{jk}+4k_2\partial_3A_i\right)\right|_{x_3=0} = 0 
\label{cond1}\end{equation}
and
\begin{equation}
\left.\epsilon_{ijk}\partial_jA_k\right|_{x_3=0} = 0\ .
\label{cond2}\end{equation}
Notice that, due to \eqref{cond2}, the $a_1$-term in the boundary action $S_{bd}$ \eqref{Sbd} vanishes. Therefore, without loss of generality, we may rule it out
\begin{equation}
a_1=0\ .
\label{a1out}\end{equation}
The equations \eqref{cond1} and \eqref{cond2} reveal an electromagnetic structure on the boundary $x_3=0$, since they suggest to define an  ``electric'' and a ``magnetic'' field:
\begin{eqnarray}
{\cal E}_i &\equiv& A_i \label{3Delectric} \\
{\cal H}_i &\equiv& k_1\epsilon_{ijk}B_{jk}+2k_2\partial_3A_i\ , \label{3Dmagnetic} 
\end{eqnarray}
which allow to identify the degrees of freedom on the boundary $x_3=0$ as the corresponding electromagnetic potentials. Indeed equations \eqref{cond1} and \eqref{cond2} are solved by introducing a 3D vector field $\xi_i(X)$ and a 3D scalar field $\Phi(X)$, respectively: 
\begin{equation}
\sqrt{M}\epsilon_{ijk}\partial_j\xi_k(X)\equiv\left.2k_1\epsilon_{ijk}B_{jk}(x)+4k_2\partial_3A_i(x)\right|_{x_3=0}
\label{defxi}\end{equation}
\begin{equation}
\frac{1}{\sqrt{M}}\partial_i\Phi(X)\left.\equiv A_i(x)\right|_{x_3=0}\ ,
\label{defphi}\end{equation}
where a massive scaling parameter $M$ has been introduced in order to make compatible the mass dimensions \eqref{dimensions} of the 4D fields $A_\mu$ and $B_{\mu\nu}$ with those of their 3D boundary counterparts $\Phi$ and $\xi_i$ \cite{Blasi:2005bk}. In 3D spacetime dimensions, in fact, vector fields and scalar fields should have the following mass dimensions
\begin{equation}
[\xi]=[\Phi]=\frac{1}{2}\ .
\label{3Ddimensions}\end{equation}\\

A comment is in order. It turns out that a particular solution of the general boundary conditions \eqref{bcA} and \eqref{bcB}  exists, which makes the physics independent of the Maxwell term. In fact, the choice
\begin{equation}
\ a_2 \neq 4k_2\ ; a_3=2k_1\ ; m= \mbox{any} 
\label{particularchoicebc}\end{equation}
implies, from \eqref{bcA} and \eqref{bcB}, (Dirichlet and) Neumann boundary conditions for the gauge field $A_i$ on $x_3=0$
\begin{equation}
\left.\partial_3A_i=0\right|_{x_3=0}\ .
\label{neumann}\end{equation}
Consequently, the Ward identities do not depend on the coupling $k_2$ of the Maxwell term in the action \eqref{Sbulk}, and the theory is indistinguishable, under any respect, from the pure topological BF theory with planar boundary. The fact that  Neumann condition for the gauge field $A_i$, which is a solution of the general boundary conditions \eqref{bcA} and \eqref{bcB}, make the Maxwell term transparent and the non-topological theory equivalent to a topological one, is the first nontrivial result of this paper. Since the aim of this paper is to study if and how the non-topological Maxwell term has an impact on the physics on the boundary, we proceed from now disregarding the solution \eqref{particularchoicebc}. In particular the boundary condition \eqref{bcB} is solved by 
\begin{equation}
a_3=0\ .
\label{a3out}\end{equation}\\

The $k_2$-Maxwell term manifests itself on the r.h.s. of the Ward identity \eqref{wardA} by means of  $\left.\partial_3 A_i\right|_{x_3=0}$. We observe that, on the boundary $x_3=0$, the fields $\left.A_i\right|_{x_3=0}$ and $\left.\partial_3 A_i\right|_{x_3=0}$ must be treated as  independent dynamical fields \cite{Karabali:2015epa}. Consequently, we need to couple, on the boundary $x_3=0$, an external source $\widehat{J}_i$  to $\left.\partial_3 A_i\right|_{x_3=0}$, as done  in \cite{Maggiore:2019wie}, where the 3D Maxwell theory with boundary has been studied: 
\begin{equation}
S_{ext}\rightarrow \hat{S}_{ext}=\int d^4x\, 
\left[\theta(x_3)\left(J_iA_i+\frac{1}{2}J_{ij}B_{ij}\right) 
+ \delta(x_3)\widehat{J}_i\partial_3A_i\right]\ .
\label{Sexthat}\end{equation}

\section{The boundary algebra}\label{sect3}

Differentiating the two Ward identities \eqref{wardA} and \eqref{wardB} with respect to the external sources $J_i(x)$, $J_{ij}(x)$ and $\hat{J}_i(x)$ and then going at $J=0$ lead to six algebraic relations. We consider the subalgebra obtained as follows. \\

Differentiating the Ward identity \eqref{wardA} with respect to $J_m(x')$, and then going at vanishing external source, we get
\begin{equation}
\partial_m^X\delta^{(3)}(X-X') =
\partial_i^X \left(
-2k_1\epsilon_{ijk}\Delta_{A_mB_{jk}}(X',X) -4k_2 \Delta_{A_m\partial_3A_i}(X',X)
\right).
\label{3.1}\end{equation}
In \eqref{3.1} $\partial_i^X\equiv\frac{\partial}{\partial X_i}$, and the time-ordered propagator between two generic fields $\Phi(X)$ and $\Phi(X)$ is defined on the generating functional of the connected Green functions $Z_c[J]$, as usual, as
\begin{equation}
\begin{split}
\Delta_{\Phi\Psi}(X,X')
&\equiv
\left. 
\frac{\delta^{(2)} Z_c}{\delta J_\Phi(X)\delta J_\Psi(X')}
\right|_{J_\Phi=J_\Psi=0}\\
&= \theta(x_0-x'_0)\langle \Phi(X)\Psi(X')\rangle +  \theta(x'_0-x_0)\langle \Psi(X')\Phi(X)\rangle.
\end{split}\label{defprop}\end{equation}
Hence, from \eqref{3.1} we have
\begin{equation}
\begin{split}
\partial^X_m\delta^{(3)}(X-X') =\ &
\delta(x_0-x'_0) \langle[A_m(X'),2k_1\widetilde{B}(X)+4k_2\partial_3A_0(X)]\rangle\\
&-2\theta(x_0-x_0')\langle\left[\partial_i^X\left(k_1\epsilon_{ijk}B_{jk}+2k_2\partial_3A_i\right)(X) A_m(X')\right]\rangle \\
&-2\theta(x_0'-x_0)\langle\left[A_m(X')\partial_i^X\left(k_1\epsilon_{ijk}B_{jk}+2k_2\partial_3A_i\right)(X) \right]\rangle,
\end{split}
\label{3.3}\end{equation}
where we defined 
\begin{equation}
\widetilde{B}(X)\equiv\epsilon_{\alpha\beta}B_{\alpha\beta}(X)\ .
\label{Bdual}\end{equation}
The last two terms on the r.h.s. of \eqref{3.3} vanish on-shell due to \eqref{cond1} so that, at vanishing external sources, we get 
\begin{equation}
\partial_m^X\delta^{(3)}(X-X') = \delta(x_0-x'_0)\langle[A_m(X'),2k_1\widetilde{B}(X)+4k_2\partial_3A_0(X)]\rangle\ .
\label{prealgebra1}\end{equation}
Following the same steps, differentiating the Ward identity \eqref{wardA} with respect to the external sources $J_{mn}(x')$ and $\widehat{J}_m(x')$ , we get, respectively
\begin{equation}
\delta(x_0-x'_0)\langle[B_{mn}(X'),k_1\widetilde{B}(X)+2k_2\partial_3A_0(X)]\rangle=0
\label{prealgebra2}\end{equation}
and
\begin{equation}
\delta(x_0-x'_0)\langle[k_1\widetilde{B}(X)+2k_2\partial_3A_0(X),\partial_3A_m(X')]\rangle=0\ .
\label{prealgebra3}\end{equation}
On the other hand, deriving the Ward identity \eqref{wardB} with respect to $J_{mn}(x')$, we obtain 
\begin{equation}
\delta(x_0-x'_0)\langle[B_{mn}(X'),k_1\widetilde{B}(X)+2k_2\partial_3A_0(X)]\rangle=0\ .
\label{prealgebra4}\end{equation}
From the above algebraic relations, we get the following subalgebra formed by equal-time commutators: 
\begin{equation}
\langle[A_\alpha(X),2k_1\widetilde{B}(X')+4k_2\partial_3A_0(X')]\rangle_{x_0=x'_0} = \partial_\alpha^X\delta^{(2)}(X-X')
\label{algebra1}\end{equation}
\begin{equation}
\langle\left[A_\alpha(X),A_\beta(X')\right]\rangle_{x_0=x'_0} = 0 
\label{algebra2} \end{equation}
\begin{equation}
\left\langle\left[2k_1\widetilde{B}(X)+4k_2\partial_3A_0(X),2k_1\widetilde{B}(X')+4k_2\partial_3A_0(X')\right]\right\rangle_{x_0=x'_0} = 0\ ,
\label{algebra3}\end{equation}
which, written in terms of the 3D  boundary fields $\xi_i(X)$ \eqref{defxi} and $\Phi(X)$ \eqref{defphi}, implies 
\begin{eqnarray}
\langle[\Phi(X),\epsilon_{\alpha\beta}\partial_\alpha\xi_\beta(X')]\rangle_{x_0=x'_0} &=& \delta^{(2)}(X-X') \label{cancomm1}\\
\langle\left[\Phi(X),\Phi(X')\right]\rangle_{x_0=x'_0} &=& 0 \label{cancomm2}\\
\langle\left[\epsilon_{\alpha\beta}\partial_\alpha\xi_\beta(X),\epsilon_{\alpha\beta}\partial_\alpha\xi_\beta(X')\right]\rangle_{x_0=x'_0} &=& 0\ . \label{cancomm3}
\end{eqnarray}

\section{The action induced on the 3D boundary}\label{sect4}

The commutators \eqref{cancomm1}, \eqref{cancomm2} and \eqref{cancomm3} can be interpreted as equal-time canonical commutation relations for the 3D canonically conjugate variables
\begin{equation}
\begin{split}
q(X) &\equiv \frac{1}{\sqrt{M}}\Phi(X)\\
p(X) &\equiv \sqrt{M}\epsilon_{\alpha\beta}\partial_\alpha\xi_\beta(X)\ ,
\end{split}
\label{qp}\end{equation}
and this allows us to identify the most general action $S_{3D}[\Phi,\xi]$ induced on the planar boundary $x_3=0$ of the 4D Maxwell-BF theory, which must display the following features:
\begin{enumerate}
\item
The 3D action $S_{3D}[\Phi,\xi]$ must be a local integrated functional of the 3D fields \eqref{defxi} and \eqref{defphi}, with canonical mass dimension equal to three.
\item
The 3D Lagrangian ${\cal L}_{3D}[\Phi,\xi]$ must be such that the relation between the canonically conjugate variables \eqref{qp} holds true: 
\begin{equation}
p(X)=\frac{\partial{\cal L}_{3D}}{\partial\dot{q}(X)}\ .
\label{pqdot}\end{equation}
This implies that the Lagrangian ${\cal L}_{3D}[\Phi,\xi]$ must contain time derivatives only in the term 
$p\dot{q}$.
\item 
The action $S_{3D}[\Phi,\xi]$ must display the two symmetries which leave invariant the definitions  \eqref{defxi} and \eqref{defphi}:
\begin{enumerate}
\item gauge symmetry
\begin{equation}
\delta_{gauge}\xi_i=\partial_i\Lambda
\label{3Dgauge}\end{equation}
\item shift symmetry
\begin{equation}
\delta_{shift}\Phi=\mbox{constant}\ .
\label{3Dshift}\end{equation}
\end{enumerate}
\end{enumerate}
The most general action satisfying the above requests is
\begin{equation}
S_{3D}[\Phi,\xi] = \int d^3X \left [
c_1(\epsilon_{\alpha\beta}F_{\alpha_\beta})(\partial_0\Phi)
+c_2 F_{\alpha\beta}F_{\alpha\beta}
+c_3 \partial_\alpha\Phi\partial_\alpha\Phi
\right],
\label{3Daction}\end{equation}
where $F_{\alpha\beta}=\partial_\alpha\xi_\beta-\partial_\beta\xi_\alpha$ and $c_i,\ i=1,2,3$ are constants. From the action $S_{3D}[\Phi,\xi]$, we get the equations of motion
\begin{eqnarray}
\frac{\delta S_{3D}}{\delta\Phi} &=& 
-2\partial_\alpha (
c_1\epsilon_{\alpha\beta}\partial_0\xi_\beta
+c_3\partial_\alpha\Phi) =0 \label{eomphi} \\
\frac{\delta S_{3D}}{\delta\xi_\alpha} &=& 
2\partial_\beta (
c_1\epsilon_{\alpha\beta}\partial_0\Phi
+2c_2F_{\alpha\beta}) =0\ . \label{eomxi} 
\end{eqnarray}

\section{Holographic constraint and duality}\label{sect5}

The equations of motion \eqref{eomphi} and \eqref{eomxi} of the scalar-vector 3D action $S_{3D}[\Phi,\xi]$ \eqref{3Daction} must be compatible with the boundary conditions \eqref{bcA} and \eqref{bcB} of the 4D Maxwell-BF theory  on the planar boundary $x_3=0$. In order to make this holographic contact \cite{Amoretti:2017xto}, the boundary condition \eqref{bcA} written in terms of the boundary degrees of freedom $\xi_i$ \eqref{defxi}  and $\Phi$ \eqref{defphi}  is
\begin{equation}
\epsilon_{ijk}\partial_j\xi_k-\kappa\partial_i\Phi =0\ ,
\label{duality}\end{equation}
where, besides \eqref{a1out} and \eqref{a3out}, we chose
\begin{equation}
a_2=0\ ,
\label{a2out}\end{equation}
 and we defined the dimensionless normalized mass parameter
\begin{equation}
\kappa\equiv\frac{m}{M}\ .
\label{defkappa}\end{equation}
We recognize in \eqref{duality} the duality relation found in \cite{Amoretti:2013nv} which extracts fermionic degrees of freedom from bosonic ones \cite{Aratyn:1983bg,Aratyn:1984jz,Amoretti:2013xya}. We shall come back to this point later. Here, it appears as the unique boundary condition which relates the 4D Maxwell-BF theory with boundary and  its holographic 3D counterpart(s), as we shall explicitly show.

Notice that the three components $i=\{0,\alpha\}$ of \eqref{duality} are
\begin{eqnarray}
\epsilon_{\alpha\beta}F_{\alpha\beta}-2\kappa\partial_0\Phi &=& 0 \label{duality0} \\
\epsilon_{\alpha\beta}\partial_0\xi_\beta-\epsilon_{\alpha\beta}\partial_\beta\xi_0+\kappa\partial_\alpha\Phi &=& 0\ , \label{dualityalpha}
\end{eqnarray}
which are compatible with \eqref{eomphi} and \eqref{eomxi} if
\begin{equation}
c_2=-\frac{1}{2\kappa}c_1\ \ ;
c_3=\kappa c_1
\label{detci}\end{equation}
and if the temporal gauge choice for the 3D gauge field $\xi_i$ is imposed
\begin{equation}
\xi_0=0\ .
\label{gaugechoicexi}\end{equation}

\section{Summary of results and discussion}\label{sect6}

When a boundary is introduced in a quantum field theory, a crucial role is played by the boundary term, which in the case studied in this paper is represented by $S_{bd}$ \eqref{Sbd}, which depends by a number of constant parameters which need to be fine tuned in order to determine the holographic theory induced on the boundary. For the 4D Maxwell-BF theory, the boundary term finally reduces to 
\begin{equation}
S_{bd}=\frac{m}{2}\int d^4x\, \delta(x_3)
A_iA_i\ ,
\label{finalSbd}\end{equation}
which depends on one massive parameter $m$ only. \\

In presence of a boundary, the question naturally arises of which boundary conditions should be imposed. The procedure described in this paper leads to the following boundary condition compatible with the holographic construction:
 \begin{equation}
\left. 2k_1\epsilon_{ijk}B_{jk} + 4k_2 \partial_3A_i  - mA_i\right|_{x_3=0} = 0\ ,
\label{finalbc} \end{equation}
which involves both the $k_1$-BF and the $k_2$-Maxwell terms. Notice that it depends on one parameter only $(m)$. It is of a nonstandard type, since it does not fall into the usual Dirichlet, Neumann or Robin boundary conditions on each field, separately. On the contrary, it mixes both the fields, and the effect of the Maxwell term is to introduce a dependence on the transverse component of the gauge field with respect to the planar boundary, which is independent of the longitudinal components, and consequently must be treated as an independent dynamical variable on the boundary. As we remarked, an unexpected consequence of the boundary conditions \eqref{bcA} and \eqref{bcB}, is that they can be solved by the set of parameters \eqref{particularchoicebc}, which implies, in particular, Neumann boundary condition for the gauge field $A_\mu$. This corresponds to eliminating any dependence from $k_2$, $i.e.$ from the Maxwell term, in the physics on the boundary, represented by the current algebra and by the boundary conditions themselves.   \\

The boundary breaks all the invariances of the unbounded theory: translations, parity and gauge symmetries. The first consequence concerns the choice of the gauge conditions, implemented by the gauge fixing term $S_{gf}$ \eqref{Sgf}, which does not need to be covariant. With deeper consequences, the Ward identities describing the Ward identities are broken by boundary terms: 
\begin{equation}
\int_0^\infty dx_3\, \partial_iJ_i = 
-\partial_i\left.\left(2k_1\epsilon_{ijk}B_{jk}+4k_2\partial_3A_i\right)\right|_{x_3=0}
\label{wardA.1}\end{equation}
\begin{equation}
\int_0^\infty dx_3\, \partial_jJ_{ij}= 
4k_1\left. \epsilon_{ijk}\partial_jA_k\right|_{x_3=0}\ .
\label{wardB.1}\end{equation}  
At $J=0$, $i.e.$ going on-shell, the vanishing of the boundary breakings lead to define on $x_3=0$ an ``electric'' and a ``magnetic'' field, and the corresponding potentials: 
\begin{eqnarray}
{\cal E}_i &\equiv& A_i \propto \partial_i\Phi \label{defphi.1} \\
{\cal H}_i &\equiv& k_1\epsilon_{ijk}B_{jk}+2k_2\partial_3A_i \propto \epsilon_{ijk}\partial_j\xi_k\ , \label{defxi.1} 
\end{eqnarray}
which allow to physically interpret the contribution of the Maxwell term as a deformation of the magnetic field on the boundary.  Without Maxwell term, $i.e.$ at $k_2=0$, the fields $A_i$ and $B_{ij}$ are  interpreted as electric and magnetic fields on the boundary, respectively. Their transverse component do not enter in the game and could safely be eliminated by Neumann boundary conditions. In presence of the Maxwell term, this is no longer true for the gauge field $A$, for which both the longitudinal and transverse components are physically important: the former as electric field, the latter as deformation of the main magnetic field represented by the dual of the $B$-field. The potentials corresponding to the boundary electric and magnetic fields  \eqref{defphi.1} and \eqref{defxi.1} are the degrees of freedom by means of which the holographic 3D theory is constructed: a vector field $\xi_i$ and a scalar field $\Phi$. 
\\

The algebra obtained from the broken Ward identities \eqref{wardA.1} and \eqref{wardB.1} by differentiating them with respect to the external sources $J$, written in terms of the boundary degrees of freedom is
\begin{eqnarray}
\langle[\Phi(X),\epsilon_{\alpha\beta}\partial_\alpha\xi_\beta(X')]\rangle_{x_0=x'_0} &=& \delta^{(2)}(X-X') \label{cancomm1.1}\\
\langle\left[\Phi(X),\Phi(X')\right]\rangle_{x_0=x'_0} &=& 0 \label{cancomm2.1}\\
\langle\left[\epsilon_{\alpha\beta}\partial_\alpha\xi_\beta(X),\epsilon_{\alpha\beta}\partial_\alpha\xi_\beta(X')\right]\rangle_{x_0=x'_0} &=& 0\ , \label{cancomm3.1}
\end{eqnarray}
which can be seen as equal-time canonical commutation relations between canonical variables $q(X)$ and $p(X)$. \\

Once identified the canonical variables, the corresponding action is found to be
\begin{equation}
S_{3D}[\Phi,\xi] = c_1\int d^3X \left [
(\epsilon_{\alpha\beta}F_{\alpha\beta})(\partial_0\Phi)
-\frac{1}{2\kappa} F_{\alpha\beta}F_{\alpha\beta}
+\kappa \partial_\alpha\Phi\partial_\alpha\Phi
\right],
\label{final3Daction}\end{equation}
which is the most general 3D local integrated functional built with a scalar and a vector field, respecting power counting and invariant under the gauge \eqref{3Dgauge} and shift \eqref{3Dshift} symmetries, and, most important, whose equations of motion are compatible with the boundary conditions \eqref{finalbc}:
\begin{equation}
\epsilon_{ijk}F_{jk}-2\kappa\partial_i\Phi =0\ .
\label{duality1}\end{equation}
The equation \eqref{duality1} coincides with the duality relation between a scalar and a vector field which has been invoked in \cite{Aratyn:1983bg,Aratyn:1984jz,Amoretti:2013nv} as the main tool for the mechanism of fermionization of bosonic degrees of freedom. In other words, the effective dynamical variables living on the boundary of the 4D Maxwell-BF are fermionic. This feature is crucial for the interpretation of the boundary degrees of freedom as the edge states of the 3D topological insulators \cite{Cho:2010rk,Blasi:2011pf}. The new fact that we are recovering here, is that this property, which has been related to the topological character of the bulk theory \cite{Amoretti:2014kba}, holds indeed also for a non-topological theory like Maxwell-BF theory. We shall comment in more detail this point later. The duality relation \eqref{duality1} has also more field theoretical consequences. Thanks to \eqref{duality1}, indeed, the action \eqref{final3Daction} is covariant, despite the appearances. In fact,  the scalar field $\Phi$ can be eliminated from the action through \eqref{duality1}, and we find the 3D Maxwell theory 
\begin{equation}
S_{Max}[\xi]= \frac{c_1}{2\kappa} \int d^3X\ F_{ij}F_{ij}\ .
\label{3Dmax}\end{equation}
Alternatively, the duality relation \eqref{duality1} allows us to trade the vector for the scalar field
\begin{equation}
S_{KK}[\Phi]= \kappa c_1\int d^3X\ \partial_i\Phi\partial_i\Phi\ .
\label{3Dkk}\end{equation}
The scalar-vector 3D theory described by \eqref{final3Daction}, the Maxwell action \eqref{3Dmax} and the massless Klein-Gordon  action \eqref{3Dkk} are all holographic counterparts of the 4D Maxwell-BF theory with planar boundary. 
The two actions \eqref{3Dmax} and \eqref{3Dkk} are related by the duality relation \eqref{duality}, which may be written in a way to emphasize its strong-weak coupling aspect:
\begin{equation}
\partial_i\Phi\leftrightarrow\epsilon_{ijk}F_{jk}\ \  
\cup
\ \  \kappa\leftrightarrow\frac{1}{\kappa}\ .
\label{kappaduality}\end{equation}
In this form, it is apparent that the coupling $\kappa$ \eqref{defkappa}  governs the regimes where one type of action dominates with respect to the other: at strong coupling (very large $\kappa$) the dominating  term is the Maxwell one, while at weak coupling (very small $\kappa$), it is the massless scalar action which dominates. At intermediate regimes, both the degrees of freedom are present, and the relevant action is \eqref{final3Daction}, whose degrees of freedom are fermionic, due to \eqref{duality1}, as we already remarked. It is this intermediate regime which is relevant for the topological insulators, for which the same action \eqref{final3Daction} has been proposed in \cite{Cho:2010rk}. Surprisingly, the order parameter $\kappa$  which distinguishes the various regimes is directly related through \eqref{defkappa} to the only effective parameter $m$ which survives in the boundary term \eqref{finalSbd}, which hence plays a much more crucial role than one might expect at first sight. \\

We conclude this paper with a remark concerning the effect of the presence of the Maxwell term in the bulk action \eqref{Sbulk} together with a general consideration on the topological nature of quantum field theories. The main feature of topological quantum field theories is the lack of local observables, the only observables being global geometrical properties of the manifolds on which they are built. Technically, this means that the local cohomology of the BRS operator is empty. Examples of topological quantum field theories are the 3D Chern-Simons theory and BF theories, which may be defined in any spacetime dimensions. The 4D Maxwell-BF theory studied in this paper is not topological, since its action depends on the metric through the Maxwell term, and its cohomological structure is nontrivial. Nevertheless the bulk action has no local observables since the equations of motion enforce the vanishing of the $F_{\mu\nu}$ tensor; thus we are looking at a different class of ``topological'' theories where the boundary is expected to carry all the physical information.  If a boundary is introduced like we did, this nontriviality  reflects in the algebra derived from the Ward identities \eqref{wardA.1} and \eqref{wardB.1}, which is rather complicated. We wrote down the four relations (out of six) \eqref{prealgebra1}, \eqref{prealgebra2}, \eqref{prealgebra3} and \eqref{prealgebra4}, where the dependence on the Maxwell term is highlighted by the coefficient $k_2$. In addition, the boundary condition \eqref{finalbc} explicitly depends on the Maxwell term, which physically results in a perturbation of the magnetic field ${\cal H}$ \eqref{defxi.1}. On the other hand, when constructing the holographic scalar-vector 3D theory, the presence of the Maxwell term is buried in the definition of the vector potential $\xi_i$ \eqref{defxi.1}, and both the 3D actions (in their duality-related representations \eqref{final3Daction}, \eqref{3Dmax} and \eqref{3Dkk} and the duality relation \eqref{duality1} are the same as in the case of pure BF with boundary. In other words, from the holographic point of view, Maxwell-BF theory with boundary is indistinguishable from the pure topological BF theory, like if holography would protect the topological character of bulk theories. 


\vspace{1cm}

%


\end{document}